\documentstyle[twoside,fleqn,espcrc2,psfig]{article}

\newcommand{\AmS}{{\protect\the\textfont2
  A\kern-.1667em\lower.5ex\hbox{M}\kern-.125emS}}

\title{\vspace{-3cm}\hspace{11.4cm}\normalsize{FERMILAB-CONF-97/256-E}\\ \vspace{2cm}
       Production of Jets at the Tevatron\thanks{
                          presented at the $5^{th}$ Topical Seminar on
                          {\it The Irresistible Rise of the Standard Model,}
                          San Miniato al Todesco, Italy, 21-25 April, 1997.}
}
\author{T. Heuring\address{Department of Physics,\\
                           Florida State University,\\
                           Tallahassee, FL 32306-3019}
                  \thanks{for the D\O\ Collaboration 
                          }}

\begin{document}

\begin{abstract}
Measurements of the inclusive jet cross section and the dijet angular
distribution using data from the Tevatron are presented.  Comparisons to NLO QCD
show good agreement below 250 GeV, but CDF data show an excess at
higher $E_T$; qualitative agreement is seen between the CDF and D\O\ cross
sections.  Analysis of the dijet angular distributions exclude quark
compositeness below 2.1 TeV.
\end{abstract}

\maketitle

\section{Introduction}

Since the discovery of the internal structure of hadrons and their subsequent
association with quarks, the next natural question to be asked is whether these
point--like particles have any substructure.  
Measuring the inclusive jet cross section and studying the angular distribution
in dijet events at the Tevatron, currently the highest energy hadron accelerator
colliding protons and antiprotons at a center--of--mass energy of 1.8 TeV,
explores matter at the smallest distance scales available.  Both collider
detector experiments at the Tevatron, CDF and D\O, have measured these
distributions to test perturbative quantum chromodynamics (QCD) and to search
for structure on the scale of $10^{-17}$ cm.

\section{Inclusive Jet Cross Section}

The inclusive jet cross section is defined as follows:
\begin{equation}
\frac{1}{\Delta \eta} \int d\eta \frac{d^2\sigma}{dE_Td\eta} =
\frac{N}{\Delta E_T \Delta \eta \int  {\cal L}dt}
\end{equation}
where $\Delta E_T$ is the bin width in transverse energy, $E_T$, $\Delta\eta$
is the pseudorapidity range covered by the measurement ($\eta = -\ln
\tan(\theta/2)$, $N$ is the number of jets in the bin, and 
$\cal L$ is the integrated luminosity.  For CDF, the rapidity interval covers
$0.1 \leq |\eta| \leq 0.7$ while D\O\ accepts jets in the region 
$|\eta| \leq 0.5$.  Both experiments correct the raw
distributions for detector effects and remove background.  In the case of
CDF, the raw distribution is corrected for energy response and resolution using
a detailed fragmentation model coupled with detector response from single
particles\cite{cdf 1a}.  The response corrections are typically $\sim 10\%$
over the whole $E_T$ range.  The resolution correction is $80\%$ at the lowest
$E_T$ ($15$ GeV $\leq E_T \leq 40$ GeV), flattening out at $\sim 10\%$ at
moderate $E_T$'s ($E_T \sim 40$ GeV) 
and increasing to $60\%$ for $E_T = 400$ GeV.  D\O\ corrects for response on a
jet--by--jet basis using a correction derived using transverse momentum
conservation in jet--jet and photon--jet events\cite{D0 escale}.  The
correction factors range from $\sim 15\%$ at low $E_T$'s to $\sim 12\%$ at the
highest $E_T$'s.  The detector resolution effects are removed by an unsmearing
procedure where an ansatz function, smeared with measured detector resolution
factors, is fitted to the raw distribution.  Comparing the smeared and
unsmeared distributions yields a correction factor\cite{D0 resolution}.  This
correction ranges from $30\%$ at low $E_T$ flattening out at $10\%$ above 100
GeV.

\begin{figure}[htb]
 \centerline{\hbox{
 \psfig{figure=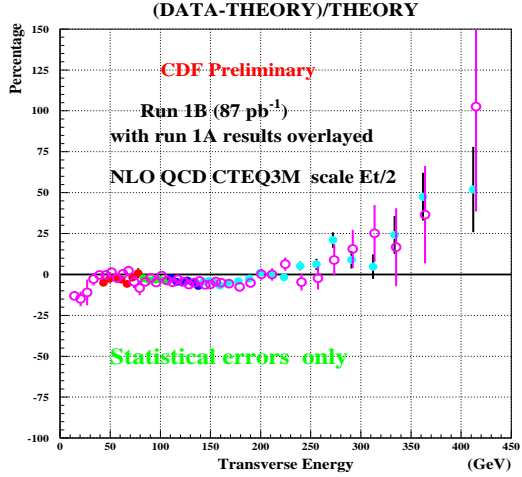,width=75 mm,height=70 mm}}}
 \caption{Results of the CDF inclusive cross section measurement from both the
 1992-93 and 1994-95 data sets compared to NLO QCD using the CTEQ3M parton
 distribution function and $\mu = E_T^{jet}/2$.}
 \label{fig:CDF inc}
\end{figure}

CDF has previously measured the inclusive jet cross section using 19 pb$^{-1}$
from the 1992-93 run at the Tevatron\cite{cdf 1a}.  These data demonstrate good
agreement with next-to-leading order (NLO) QCD predictions below 250 GeV but
showed an excess above this.  The higher statistics available from the 1994-95
run exhibit a similar excess.  Figure \ref{fig:CDF inc} shows both data sets
compared to NLO QCD produced by the EKS program\cite{eks} using CTEQ3M parton
distribution functions\cite{cteq3m} with the renormalization and
factorization scales ($\mu$) set equal to  the $E_T$ of each jet divided by two.
Although the systematic errors for the 1994-95 run are still being determined,
they are expected to be of the same magnitude as was found on the 1992-93 data.

\begin{figure}[htb]
 \centerline{\hbox{
 \psfig{figure=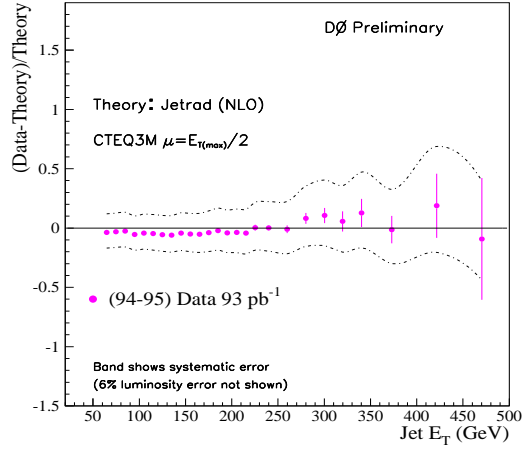,width=75 mm,height=70 mm}}}
 \caption{Results of the D\O\ inclusive cross section measurement from the
 1994-95 data set compared to NLO QCD using the CTEQ3M parton distribution
 function and $\mu = E_T^{max}/2$.}
 \label{fig:D0 inc}
\end{figure}

The results from D\O\ are shown in Fig. \ref{fig:D0 inc}.  In this case the data
are compared with NLO QCD from the JETRAD program\cite{jetrad} with $\mu =
E_T^{max}/2$, the $E_T$ of the highest $E_T$ jet in each event, again using the
CTEQ3M parton distribution function.  The statistical errors are indicated on
each point with the systematic error indicated by the band.  The $6\%$
luminosity uncertainty is not included.  The NLO QCD predictions are in very
good agreement with the data over the entire $E_T$ range.

\begin{figure}[htb]
 \centerline{\hbox{
 \psfig{figure=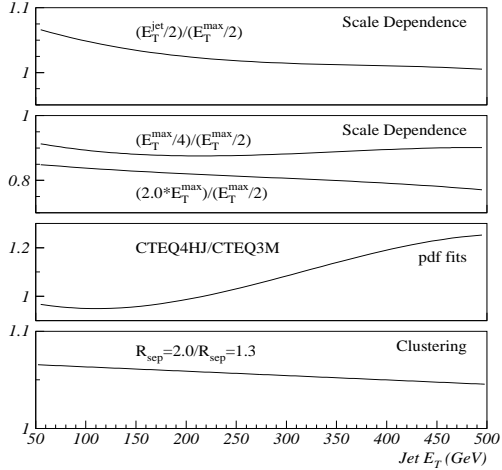,width=75 mm,height=70 mm}}}
 \caption{Effects of different theoretical parameters in the inclusive jet cross
 section.  Comparisons are made to a standard calculations where $|\eta| \leq 
 0.5$, $\mu = E_T^{max}/2$, and $R_{sep} = 1.3R$ using the CTEQ3M parton
 distribution function.}
 \label{fig:inc theory}
\end{figure}

The difficulty in comparing the CDF and D\O\ results in Figs. \ref{fig:CDF inc}
and \ref{fig:D0 inc} arises due to the many different choices available
when comparing to theory.  Figure \ref{fig:inc theory} illustrates the
dependence of the 
theory predictions to some of these choices.  Aside from the
obvious multiplicative factors typically used to scale the choice of $\mu$
which can result in $15\%$ shift in the theoretical prediction, choosing $E_T$
of the individual jets or the maximum $E_T$ in each event as the scale can
result in a $5\%$ shift.  Choices in parton distribution functions, some of
which allow for a significant increase in the production of large $E_T$
jets\cite{cteqhj}, can result in $20\%$ effects.  Although both experiments use
a cone algorithm with a radius, $R = \sqrt{\Delta\eta^2 + \Delta\phi^2}$, of 
0.7, CDF uses the standard Snowmass parameters\cite{snowmass}.  In this case,
two partons can be separated by as much as twice the cone radius, $R_{sep} =
2R$.  D\O, on the other hand, sets $R_{sep} = 1.3R$ to better match jet
clustering effects observed in the D\O\ data.  Qualitative comparisons of the
two measurements are in good agreement; a more quantitative analysis is
underway.

%In an attempt to eliminate the effects caused by the different $\eta$ regions
%used by each experiment and to reduce the effect of different theory choices,
%D\O\ has repeated their analysis using the CDF fiducial cut of 
%$0.1 \leq |\eta| \leq 0.7$.  The results were fitted to a smooth function and
%compared to the CDF results in Fig. \ref{fig:CDF comp}.  The two experimental
%results appear to agree qualitatively, the CDF points lying within the
%D\O\ error band, but more quantitative comparisons are underway.

%\begin{figure}[htb]
% \centerline{\hbox{
% \psfig{figure=cdf_comp.eps,width=75 mm,height=70 mm}}}
% \caption{Comparison of the CDF and D\O\ results.  The D\O\ analysis was
% repeated using the CDF experimental cuts and fit to a smooth curve.  The
% results indicate that the two experiments are in good qualitative agreement.}
% \label{fig:CDF comp}
%\end{figure}

\section{Dijet Angular Distribution}

Although the inclusive jet cross section is one way to search for quark
substructure, the measurement suffers from sensitivity to the input parton
distribution function choice.  An alternative to this approach that is less
sensitive to this problem is to measure the angular distribution of the two
leading jets in dijet events.  This measurement is analogous to the original
Rutherford scattering experiments from the turn of the century.  Evidence of
substructure will manifest itself as distributions which are more isotropic
than those expected from the interaction of point--like quarks.

To look for quark compositeness, comparisons will be made between the data and
QCD predictions incorporating compositeness models.  To date, NLO QCD
predictions including compositeness are not available.  Therefore, the
\begin{figure}[htb]
 \centerline{\hbox{
 \psfig{figure=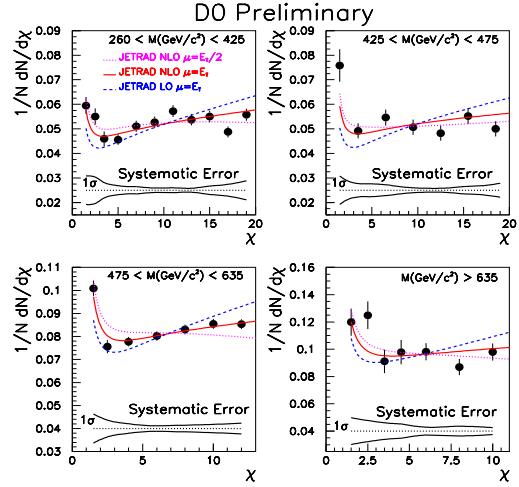,width=75 mm,height=70 mm}}}
 \caption{The dijet angular distribution from D\O\ in four mass bins are
 compared to NLO QCD with $\mu = E_T^{max}$ and $\mu = E_T^{max}/2$ and to LO
 QCD with $\mu = E_T^{max}$.  Note that it is possible to distinguish between
 LO and NLO QCD as well as different choices of $\mu$.  % at large values of
 %$\chi$.
 The errors are statistical only with the systematic error indicated by the
 band.}
 \label{fig:D0 dijet}
\end{figure}
effects\cite{lane} are determined at LO, comparing predictions with and without
such features.  The NLO theory is then scaled by the ratio of the two LO
theories. The measurement will be made in terms of the variable $\chi$,
    \begin{equation}
    \chi \equiv e^{(|\eta_1-\eta_2|)} =
    \frac{1+\cos{\theta^*}}{1-\cos{\theta^*}}
    \end{equation}
where $\eta_{1,2}$ are the pseudorapidities of the 
two leading jets and $\theta^*$ is
the center--of--mass scattering angle.  This has the virtue of flattening the
angular distributions making comparisons to theory easier.

Figure \ref{fig:D0 dijet} shows the dijet angular distributions from D\O\ for
four different mass bins.  Statistical errors bars are included on the
individual points and the systematic error is indicated by the bands.  The data
is compared with NLO QCD predictions from JETRAD using two different $\mu$
\begin{figure}[htb]
 \centerline{\hbox{
 \psfig{figure=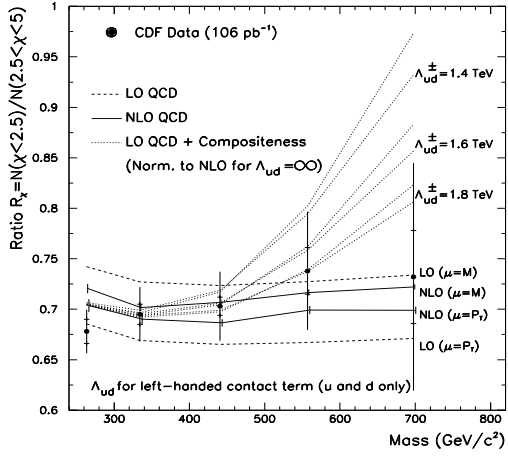,width=75 mm,height=70 mm}}}
 \caption{$R_\chi$ vs. mass from the CDF analysis compared to LO and NLO QCD as
 well as models with various compositeness scales.  The inner error bars
 indicate the statistical error while the outer include the statistical and
 systematic added in quadrature.}
 \label{fig:CDF dijet}
\end{figure}
scales, $E_T^{max}$ and $E_T^{max}/2$.  Since the angular distribution
predictions are sensitive to the choice of scale, any compositeness limit will
have to take this into account.  In addition, the extended $\chi$ range
available at D\O\ also allows some discrimination between LO and NLO theoretical
predictions.

Finally, to search for compositeness we define a variable $R_\chi$,
    \begin{equation}
    R_\chi = \frac{\# \:of\: events\: \chi<\chi_0}
                  {\# \:of\: events\: \chi>\chi_0}
    \end{equation}
This compares the number of events in a region where compositeness effects
should be minimal to a region where they should be enhanced.  The measurements
are made in mass bins with $\chi_0 = 2.5$ for CDF and 4 for D\O.  By
comparing the measured ratio to various QCD compositeness predictions, a
compositeness limit can be extracted.

The results of the CDF analysis are shown in Fig. \ref{fig:CDF dijet}.  Here
\begin{figure}[htb]
 \centerline{\hbox{
 \psfig{figure=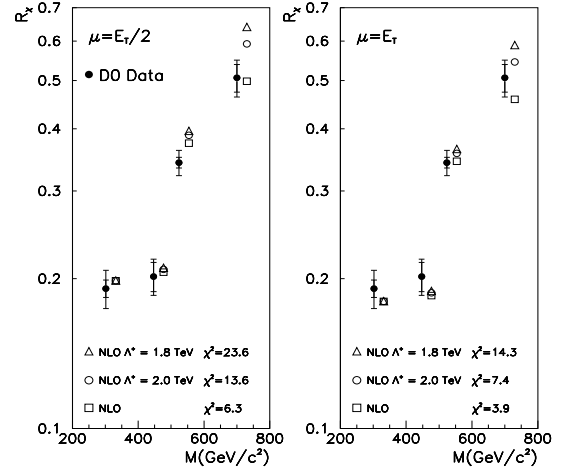,width=75 mm,height=70 mm}}}
 \caption{$R_\chi$ vs. mass from the D\O\ analysis compared NLO QCD with and
 without compositeness.  The inner errors represent the statistical error while
 the outer contain the statistical and systematic added in quadrature.  The
 $\chi^2$ values for 4 d.o.f. are shown for different $\Lambda$ values.}
 \label{fig:D0 dijet limit}
\end{figure}
jets were required to satisfy $0.1 \leq |\eta| \leq 2.0$ and $\chi \leq 5$. 
Also shown are various theoretical predictions showing the effects of
LO vs. NLO QCD, the effects of different renormalization scales, and the effect
of different compositeness terms.  For models where all quarks are composite,
CDF excludes at the $95\%$ confidence 
level regions with $\Lambda^+ \leq 1.8$
TeV and $\Lambda^- \leq 1.6$ TeV.

The D\O\ compositeness limit results are shown in Fig. \ref{fig:D0 dijet limit}.
Jets are accepted out to $|\eta| \leq 3.0$, extending the $\chi$ reach to 20
when kinematically accessible.  The data points are compared to NLO QCD models
with various compositeness scales and different $\mu$ scales.  For models where
all quarks are composite, D\O\ excludes at the $95\%$ confidence level regions
with $\Lambda^+ \leq 2.3$ TeV for $\mu = E_T^{max}/2$ and $\Lambda^+ \leq 2.1$
TeV for $\mu = E_T^{max}$.

\section{Conclusion}

Measurements of the inclusive jet cross section and the dijet angular
distribution have been made by both the CDF and D\O\ experiments.  While the
excess that was reported in the earlier CDF inclusive jet analysis seems to
persist in the new data set, the D\O\ data appear to agree with NLO QCD over
the entire $E_T$ range.  No evidence of quark substructure was evident in the
dijet angular distribution.


\begin{thebibliography}{9}

\bibitem{cdf 1a} F.~Abe {\it et al.}, CDF Collab., Phys. Rev. Lett. {\bf 77},
        438 (1996).

\bibitem{D0 escale} R. Kehoe, Proceedings of 6th International Conference on
        Calorimetry in High-energy Physics, 1996. pp. 349-358. 

\bibitem{D0 resolution} G. Blazey, Proceedings of 31st Rencontres de Moriond,
        1996, pp. 155-164. 


\bibitem{eks}S.~Ellis,  Z.~Kunszt, and  D.~Soper, Phys. Rev.  Lett. {\bf 64},
        2121 (1990).

\bibitem{cteq3m}J.~Botts {\it et al.}, CTEQ Collab., Phys. Rev. D {\bf 51}, 4763
        (1995).

\bibitem{jetrad}W.~Giele,  E.W.N.~Glover, and D.~Kosower,  Phys. Rev. Lett. {\bf
        73}, 2019 (1994).

\bibitem{cteqhj}J. Huston {\it et al.}, PRL77, 444(1996).

\bibitem{snowmass}John~E.~Huth {\it  et al.}, Research Directions
        for the Decade  Snowmass 1990, pp.  134-136, edited by  Edmond~L.~Berger
        (World Scientific, Singapore 1990).

\bibitem{lane} K. Lane, BUHEP-96-8, hep-ph/9605257 (1996).

\end{thebibliography}
\end{document}